\documentclass[twocolumn,aps,showpacs,floatfix,prc]{revtex4}
\usepackage{graphicx}
\usepackage[dvips]{epsfig}

\begin{document}

\title{ Gravitational waves from isolated neutron stars: \\ mass dependence of r-mode instability }

\author{ Somnath Mukhopadhyay$^{1*\S}$, Joydev Lahiri$^{2*}$, Debasis Atta$^{3\dagger\S}$, Kouser Imam$^{4\ddagger}$  and D. N. Basu$^{5*\S}$ }

\affiliation{$^*$ Variable  Energy  Cyclotron Centre, 1/AF Bidhan Nagar, Kolkata 700064, India}
\affiliation{ $^{\dagger}$ Government General Degree College, Kharagpur II, West Bengal 721149, India}
\affiliation{ $^{\ddagger}$ Department of Physics, Aliah University, IIA/27, New Town, Kolkata 700156, India}
\affiliation{$^{\S}$ Homi Bhabha National Institute, Training School Complex, Anushakti Nagar, Mumbai 400085}

\email[E-mail 1: ]{somnathm@vecc.gov.in}
\email[E-mail 2: ]{joy@vecc.gov.in}
\email[E-mail 3: ]{debasisa906@gmail.com}
\email[E-mail 4: ]{kouserblackhole@gmail.com}
\email[E-mail 5: ]{dnb@vecc.gov.in}

\date{\today }

\begin{abstract}

    In this work we study the r-mode instability windows and the gravitational wave signatures of neutron stars in the slow rotation approximation using the equation of state obtained from the density dependent M3Y effective interaction.  We consider the neutron star matter to be $\beta$-equilibrated neutron-proton-electron matter at the core with a rigid crust. The fiducial gravitational and viscous timescales, the critical frequencies and the time evolutions of the frequencies and the rates of frequency change are calculated for a range of neutron star masses. We show that the young and hot rotating neutron stars lie in the r-mode instability region. We also emphasize that if the dominant dissipative mechanism of the r-mode is the shear viscosity along the boundary layer of the crust-core interface, then the neutron stars with low $L$ value lie in the r-mode instability region and hence emit gravitational radiation. 
\vspace{0.2cm}    

\end{abstract}


\maketitle

\noindent
\section{Introduction}
\label{Section 1} 
     
    Quasinormal modes of rapidly rotating isolated and accreting compact stars act as sensitive probes for general relativistic effects such as gravitational waves and also of the properties of ultradense matter. Temporal changes in the rotational period of neutron stars (NSs) can reveal the internal changes of the stars with time. Gravitational waves from rotational instabilities of pulsars can provide an insight about the nature of the high density Equation of State (EoS). Detecting these waves by LIGO and VIRGO will provide a new pathway in the field of asteroseismology.  
    
    Rotational instabilities in NSs come in different flavours, but they have one general feature in common: they can be directly associated with unstable modes of oscillation \cite{Andersson1998,Andersson2003,Freidman1998,Provost1981,Andersson2001,Bondarescu2007}. In the present work the r-mode instability has been discussed with reference to the EoS obtained using the density dependent M3Y (DDM3Y) effective nucleon-nucleon (NN) interaction. The discovery of r-mode oscillation in neutron star (NS) by Anderson \cite{Andersson1998} and confirmed by Friedman and Morsink \cite{Freidman1998} opened the window for study of the gravitational wave emitted by NSs by using advance detecting system. Also it provides the possible explanation for the spin down mechanism in the hot young NSs as well as in spin up cold old accreting NSs.
 
    The r-mode oscillation is analogous to Rossby wave in the ocean and results from perturbation in velocity field of the star with little disturbance in the star's density. In a non-rotating star the r-modes are neutral rotational motions. In a rotating star Coriolis effects provide a weak restoring force that gives them genuine dynamics. The r-mode frequency always has different signs in the inertial and rotating frames. That is, although the modes appear retrograde in the rotating system, an observer in the inertial frame shall view them as prograde. To the leading order, the pattern speed of the mode is \cite{Andersson1999,Papaloizou1978}
    
\begin{equation}
\sigma=\frac{\left(l-1\right)\left(l+2\right)}{l\left(l+1\right)} \Omega
\label{eqn1}
\end{equation}

    Since, $0<\sigma<\Omega$ for all $l\geq 2$, where $\Omega$ is the angular velocity of the star in the inertial frame, the r-modes are destabilized by the standard Chandrasekhar-Friedman-Schutz (CFS) mechanism and are unstable because of the emission of gravitational waves. The gravitational radiation that the r-modes emit comes from their time-dependent mass currents. This is the gravitational analogue of magnetic monopole radiation. The quadrupole $l=2$ r-mode is more strongly unstable to gravitational radiation than any other mode in neutron stars. Further, these modes exist with velocity perturbation if and only if $l=m$ mode \cite{Provost1981,Andersson1999}. This emission in gravitational waves causes a growth in the mode energy $E_{rot}$ in the rotating frame, despite  decrease in the inertial-frame energy $E_{inertial}$. This puzzling effect can be understood from the relation between the two energies,
    
\begin{equation}
E_{rot}=E_{inertial}-\Omega J
\label{eqn2}
\end{equation}
where the angular momentum of the star is J. From this it is clear that $E_{rot}$ may increase if both $E_{inertial}$ and J decrease. The frequencies of these r-modes, in the lowest order terms in an expansion in terms of angular velocity $\Omega$ is \cite{Papaloizou1978,Lindblom1998}

\begin{equation}
\omega=-\frac{\left(l-1\right)\left(l+2\right)}{l+1} \Omega.
\label{eqn3}
\end{equation}

    The instability in the mode grows because of gravitational wave emission which is opposed by the viscosity \cite{Lindblom1987}. For the instability to be relevant, it must grow fast than it is damped out by the viscosity. So the time scale for gravitationally driven instability needs to be sufficiently short to the viscous damping time scale. The amplitude of r-modes evolves with time dependence $ e^{i\omega t - t/\tau } $ as a consequence of ordinary hydrodynamics and the influence of the various dissipative processes. The imaginary part of the frequency $1/\tau$ is determined by the effects of gravitational radiation, viscosity, etc \cite{Lindblom2000,Lindblom1998,Owen1998}. The time-scale associated with the different process involve the actual physical parameters of the neutron star. In computing these physical parameters the role of nuclear physics comes into picture, where one gets a platform to constrain the uncertainties existing in the nuclear EoS. The present knowledge on nuclear EoS under highly isospin asymmetric dense situation is quite uncertain. So correlating the predictions of the EoSs obtained under systematic variation of the physical properties, to the r-mode observables can be of help in constraining the uncertainity associated with the EoS.
     
\noindent
\section{Dissipative time scales and stability of the r-modes} 
\label{Section 2}
 
    The concern here is to study the evolution of the r-modes due to the competition of gravitational radiation and  dissipative influence of viscosity. For this purpose it is necessary to consider the effects of radiation on the evolution of mode energy. This is expressed as the integral of the fluid perturbation \cite{Lindblom1998,Lindblom1999},
    
\begin{equation}
\widetilde{E}=\frac{1}{2}\int{\left[ \rho \delta \vec{v}.\delta \vec{v}^{*}+\left(\frac{\delta p}{\rho}-\delta \Phi \right)\delta \rho^{*}\right]}d^{3}r,
\label{eq4}
\end{equation}
with $\rho$ being the mass density profile of the star, $\delta \vec{v}$, $\delta p$, $\delta \Phi$ and $\delta \rho $ are perturbations in the velocity, pressure, gravitational potential and density due to oscillation of the mode respectively. The dissipative time scale of an r-mode is \cite{Lindblom1998},

\begin{equation}
\frac{1}{\tau_{i}}=-\frac{1}{2\widetilde{E}}\left(\frac{d\widetilde{E}}{dt}\right)_{i},
\label{eq5}
\end{equation}
where the index `$i$' refers to the various dissipative mechanisms, i.e., gravitational wave emissions and viscosity (bulk and shear).

For the lowest order expressions for the r-mode $\delta \vec{v}$ and $\delta \rho$ the expression for energy of the mode in Eq.(4) can be reduced to a one-dimensional integral \cite{Lindblom1998,Vidana2012} 

\begin{equation}
\widetilde{E}=\frac{1}{2}\alpha_r^{2} R^{-2l+2} \Omega^{2} \int^{R}_{0} \rho(r) r^{2l+2} dr, 
\label{eq6}
\end{equation}
where R is the radius of the NS, $\alpha_r$ is the dimensionless amplitude of the mode, $\Omega$ is the angular velocity of the NS and $\rho(r)$ is the radial dependance of the mass density of NS. Since the expression of $(\frac{d\widetilde{E}}{dt})$ due to gravitational radiation \cite{Thorne1980,Owen1998} and viscosity \cite {Lindblom1991,Owen1998,Lindblom2000} are known, Eq.(5) can be used to evaluate the imaginary part $\frac{1}{\tau}$. It is convenient to decompose $\frac{1}{\tau}$ as
\begin{equation}
\frac{1}{\tau(\Omega,T)}=\frac{1}{\tau_{GR}(\Omega,T)}+\frac{1}{\tau_{BV}(\Omega,T)}+\frac{1}{\tau_{SV}(\Omega,T)},
\label{eq7}
\end{equation}
where $1/\tau_{GR}$, $1/\tau_{BV}$ and $1/\tau_{SV}$ are the contributions from gravitational radiation, bulk viscosity and shear viscosity, respectively, and are given by \cite{Owen1998,Lindblom2000}

\begin{eqnarray}
\frac{1}{\tau_{GR}}=-\frac{32 \pi G \Omega^{2l+2}}{c^{2l+3}} \frac{(l-1)^{2l}}{[(2l+1)!!]^2}\left(\frac{l+2}{l+1}\right)^{(2l+2)}\nonumber \\
 \times\int^{R_{c}}_{0}\rho(r)r^{2l+2} dr, 
\label{eq8}
\end{eqnarray}
\begin{equation}
\frac{1}{\tau_{BV}}\approx \frac{4 R^{2l-2}}{(l+1)^2} \int {\xi |{\frac{\delta \rho}{\rho}}|^{2}} d^{3}r\left(\int^{R_{c}}_{0}\rho(r)r^{2l+2} dr\right)^{-1},
\label{eq9}
\end{equation}
\begin{eqnarray}
\frac{1}{\tau_{SV}}=\left[\frac{1}{2\Omega} \frac{2^{l+3/2}(l+1)!}{l(2l+1)!!I_{l}}\sqrt{\frac{2\Omega R_{c}^{2} \rho_{c}}{\eta_c}}\right]^{-1}\nonumber \\
\times\left[\int^{R_{c}}_{0} \frac{\rho(r)}{\rho_{c}}\left(\frac{r}{R_{c}}\right)^{2l+2} \frac{dr}{R_c}\right]^{-1}, 
\label{eq10}
\end{eqnarray}
where G and c are the gravitational constant and velocity of light respectively; $\delta \rho$ in Eq.(9) is the density perturbation associated with r-modes and $\xi$ is the bulk viscosity of the fluid, $R_{c}$, $\rho_{c}$, $\eta_{c}$ in Eq.(10) are the radius, density and shear viscosity of the fluid at the outer edge of the core respectively.

    The expression for $\frac{1}{\tau_{BV}}$ in Eq.(9) is approximate and the exact expression should contain the Lagrangian density perturbation $\Delta \rho$ instead of Eulerian perturbation $\delta \rho$. The shear viscosity time scale in Eq.(10) is obtained by considering the shear dissipation in the viscous boundary layer between solid crust and the liquid core with the assumption that the crust is rigid and hence static in rotating frame \cite{Lindblom2000}.

    The motion of the crust due to mechanical coupling to the core effectively increases $\tau_{SV}$ by $(\frac{ \Delta v}{v})^{-2}$, where $\frac{\Delta v}{v}$ is the difference in the velocities in the inner edge of the crust and outer edge of the core divided by the velocity of the core \cite{Levin2001}.

    Bildsten and Ushomirsky  \cite{Bildsten2000} have first estimated this effect of solid crust on r-mode instability and shown that the shear dissipation in this viscous boundary layer decreases the viscous damping time scale by more than $10^5$ in old acreting neutron stars and more than $10^7$ in hot, young neutron stars. $I_{l}$ in Eq.(10) has the value $I_{2}=0.80411$, for $l=2$ \cite{Lindblom2000}. 

    Moreover, the bulk viscous dissipation is not significant for temperature of the star below $10^{10}$ K and in this range of temperature the shear viscosity is the dominant dissipative mechanism, We have restricted our study in this work to the range of the temperature $T<10^{10}$ K and included only shear dissipative mechanism. The studies is similar to the one done by Moustakidis \cite{Moustakidis2015}, where we have mainly examined the influence of neutron star EoS and the gravitational mass on the instability boundary and other relevant quantities, such as, critical frequency and temperature, etc. for a neutron star using the DDM3Y effective interaction \cite{BCS08}.

    As mentioned above, we have studied the instability within $T \leq 10^{10}$ K, the dominant dissipation mechanism is the shear viscosity in the boundary layer for which the time scale is given in Eq.(7), where $\eta_c$  is the viscosity of the fluid. In the temperature range $T \geq 10^9$ K, the dominant contribution to shear is from neutron-neutron (nn) scattering and below $T \leq 10^9$, it is the electron-electron (ee) scattering that contributes to shear primarily \cite{Lindblom2000}. Therefore,
    
\begin{equation}
\frac{1}{\tau_{SV}}=\frac{1}{\tau_{ee}}+\frac{1}{\tau_{nn}},
\label{eq11}
\end{equation}
where $\tau_{ee}$ and $\tau_{nn}$ can be obtained from Eq.(10) using the corresponding value of $\eta_{SV}^{ee}$ and $\eta_{SV}^{nn}$. These are given by \cite{Flowers1979,Cutler1987}

\begin{equation}
\eta_{SV}^{ee}=6 \times 10^{6} \rho^{2} T^{-2} ~~~~~({\rm g~cm^{-1}~s^{-1}}), 
\label{eq12}
\end{equation}

\begin{equation}
\eta_{SV}^{nn}=347 \rho^{9/4} T^{-2} ~~~~~({\rm g~cm^{-1}~s^{-1}}), 
\label{eq13}
\end{equation}
where all the quantities are given in CGS units and T is measured in K. In order to have transparent visualisation of the role of angular velocity and temperature on various time scales, it is useful to factor them out by defining fiducial time scales. Thus, we define fiducial shear viscous time scale $\widetilde{\tau}_{SV}$ such that \cite{Lindblom2000,Lindblom1998},

 \begin{equation}
\tau_{SV}=\widetilde{\tau}_{SV} \left(\frac{\Omega_0}{\Omega}\right)^{1/2} \left(\frac{T}{10^8 K}\right),
\label{eq14}
\end{equation}
and fiducial gravitational radiation time scale $\widetilde{\tau}_{GR}$ is defined through the relation \cite{Lindblom2000,Lindblom1998},
 
\begin{equation}
\tau_{GR}=\widetilde{\tau}_{GR} \left(\frac{\Omega_0}{\Omega}\right)^{2l+2},
\label{eq15}
\end{equation}
where $\Omega_0=\sqrt{ \pi G \bar{\rho}}$ and $\bar{\rho}= 3M/4 \pi R^3$ is the mean density of NS having mass $M$ and radius $R$.
Thus Eq.(7) (neglecting bulk viscosity contributions) becomes

\begin{equation}
\frac{1}{\tau(\Omega,T)}=\frac{1}{\widetilde{\tau}_{GR}}\left(\frac{\Omega}{\Omega_0}\right)^{2l+2}+\frac{1}{\widetilde{\tau}_{SV}}\left(\frac{\Omega}{\Omega_0}\right)^{1/2} \left(\frac{10^8 K}{T}\right).
\label{eq16}
\end{equation}

    Dissipative effects cause the mode to decay exponentially as $e^{-t/\tau}$ i.e. the mode is stable, as long as $\tau>0$. From Eq.(8) and Eq.(10) it can be seen that $\widetilde{\tau}_{SV}>0$, while $\widetilde{\tau}_{GR}<0$. Thus gravitational radiation drives these modes towards instability while viscosity tries to stabilize them. For small $\Omega$ the gravitational radiation contribution to $1/\tau$ is very small since it is proportional to $\Omega^{2l+2}$. Thus for sufficiently small angular velocity, viscosity dominates and the mode is stable. But for sufficiently large $\Omega$ gravitational radiation will dominate and drive the mode unstable. For a given temperature and mode $l$ the equation for critical angular velocity $\Omega_c$ is obtained from the condition $\frac{1}{\tau(\Omega_c,T)}=0$. At a given T and mode $l$, the equation for the critical velocity is a polynomial of order $l+1$ in $\Omega_c^{2}$ and thus each mode has its own characteristic $\Omega_c$. Since the smallest of these, i.e. $l=2$, is the dominant contributor, study is being done for this mode only. The critical angular velocity $\Omega_c$ for this mode is obtained to be
    
\begin{equation}
\left(\frac{\Omega_c}{\Omega_0}\right)=\left(-\frac{\widetilde{\tau}_{GR}}{\widetilde{\tau}_{SV}}\right)^{2/11}\left(\frac{10^8 K}{T}\right)^{2/11}.
\label{eq17}
\end{equation}

    The angular velocity of a neutron star can never exceed the Kepler velocity $\Omega_K\approx\frac{2}{3}\Omega_0$. Thus, there is a critical temperature below which the gravitational radiation is completely suppressed by viscosity. This critical temperature is given by \cite{Lindblom2000}
    
\begin{equation}
\frac{T_c}{10^8 K}=\left(\frac{\Omega_0}{\Omega_c}\right)^{11/2} \left(-\frac{\tilde{\tau}_{GR}}{\widetilde{\tau}_{SV}}\right)\approx(3/2)^{11/2}\left(-\frac{\widetilde{\tau}_{GR}}{\widetilde{\tau}_{SV}}\right).
\label{eq18}
\end{equation}
The critical angular velocity is now expressed in terms of critical temperature from Eq.(14) and Eq.(15) as

\begin{equation}
\left(\frac{\Omega_c}{\Omega_0}\right)=\frac{\Omega_K}{\Omega_0}\left( \frac{T_c}{T}\right)^{2/11}\approx(2/3)\left( \frac{T_c}{T}\right)^{2/11}.
\label{eq19}
\end{equation}
So, once the neutron star EoS is ascertained, then all physical quantities necessary for the calculation of r-mode instability can be performed.

    Further, following the work of Owen et al. \cite{Owen1998} the evolution of the angular velocity, as the angular momentum is radiated to infinity by the gravitational radiation is given by
    
\begin{equation}
\frac{d\Omega}{dt}=\frac{2\Omega}{\tau_{GR}}\frac{\alpha_r^2Q}{1-\alpha_r^2Q},
\label{eq20}
\end{equation}
where $\alpha_r$ is the dimensionless r-mode amplitude and $Q=3 \widetilde{J}/2 \widetilde{I}$ with,

\begin{equation}
\widetilde{J}=\frac{1}{MR^4}\int^{R}_{0}\rho(r)r^{6} dr
\label{eq21}
\end{equation}
and
\begin{equation}
\widetilde{I}=\frac{8\pi}{3MR^2}\int^{R}_{0}\rho(r)r^{4} dr.
\label{eq22}
\end{equation}
$\alpha_r$ is treated as free parameter whose value varies within a wide range $1-10^{-8}$. Under the ideal consideration that the heat generated by the shear viscosity is same as that taken out by the emission of neutrinos \cite{Bondarescu2009,Moustakidis2015}, Eq.(20) can be solved for the angular frequency $\Omega(t)$ as

\begin{equation}
 \Omega(t)=\left(\Omega^{-6}_{in}-\textsl{C}t\right)^{-1/6},
\label{eq23}
\end{equation}
where 

\begin{equation}
 \textsl{C}=\frac{12\alpha_r^2Q}{\widetilde{\tau}_{GR}\left(1-\alpha_r^2Q\right)}\frac{1}{\Omega_0^6},
\label{eq24}
\end{equation}
and $\Omega_{in}$ is considered as a free parameter whose value corresponds to be the initial angular velocity. The spin down rate can be obtained from Eq.(20) to be,

\begin{equation}
\frac{d\Omega}{dt}=\frac{\textsl{C}}{6}\left(\Omega^{-6}_{in}-\textsl{C}t\right)^{-7/6}.
\label{eq25}
\end{equation}

    The neutron star spin shall decrease continually until it approaches its critical angular velocity $\Omega_c$. The time $t_c$ taken by neutron star to evolve from its initial value $\Omega_{in}$ to its minimum value  $\Omega_{c}$ is given by
\begin{equation}
t_c=\frac{1}{\textsl{C}}\left(\Omega_{in}^{-6}-\Omega_{c}^{-6}\right).
\label{eq26}
\end{equation}
\noindent
\section{Nuclear Equation of State} 
\label{Section 3}

    The EoS for nuclear matter is obtained by using the isoscalar and the isovector \cite{Sa83} components of M3Y effective NN interaction along with its density dependence. The nuclear matter calculation is then performed which enables complete determination of this density dependence. The minimization of energy per nucleon determines the equilibrium density of the symmetric nuclear matter (SNM). The variation of the zero range potential with energy, over the entire range of the energy per nucleon $\epsilon$, is treated properly by allowing it to vary freely with the kinetic energy part $\epsilon^{kin}$ of $\epsilon$. This treatment is more plausible as well as provides excellent result for the SNM incompressibility $K_\infty$. Moreover, the EoS for SNM is not plagued with the superluminosity problem. 
        
    The energy per nucleon $\epsilon$ for isospin asymmetric nuclear matter (IANM) can be derived within a Fermi gas model of interacting neutrons and protons as \cite{BCS08} 
    
\begin{equation}
 \epsilon(\rho,X) = [\frac{3\hbar^2k_F^2}{10m}] F(X) + (\frac{\rho J_v C}{2}) (1 - \beta\rho^n)  
\label{eq27}
\end{equation}
\noindent
where isospin asymmetry $X= \frac{\rho_n - \rho_p} {\rho_n + \rho_p}$, $\rho = \rho_n+\rho_p$ with $\rho_n$, $\rho_p$ and $\rho$ being the neutron, proton and nucleonic densities respectively, $m$ is the nucleonic mass, $k_F$=$(1.5\pi^2\rho)^{\frac{1}{3}}$ which equals Fermi momentum in case of SNM, $\epsilon^{kin}=[\frac{3\hbar^2k_F^2}{10m}] F(X)$ with $F(X)$=$[\frac{(1+X)^{5/3} + (1-X)^{5/3}}{2}]$ and $J_v$=$J_{v00} + X^2 J_{v01}$, $J_{v00}$ and $J_{v01}$ represent the volume integrals of the isoscalar and the isovector parts of the M3Y interaction. The isoscalar and isovector components $t_{00}^{M3Y}$ and $t_{01}^{M3Y}$ of the M3Y effective NN interaction are given by $t_{00}^{M3Y}(s, \epsilon)=7999\frac{\exp( - 4s)}{4s}-2134\frac{\exp( - 2.5s)}{2.5s}+J_{00}(1-\alpha\epsilon)\delta(s)$ and $t_{01}^{M3Y}(s, \epsilon)=-4886\frac{\exp( - 4s)}{4s}+1176\frac{\exp( - 2.5s)}{2.5s}+J_{01}(1-\alpha\epsilon)\delta(s)$, respectively, with $J_{00}$=-276 MeVfm$^3$, $J_{01}$=228 MeVfm$^3$, $\alpha=0.005$MeV$^{-1}$. The DDM3Y effective NN interaction is given by $v_{0i}(s,\rho, \epsilon) = t_{0i}^{M3Y}(s, \epsilon) g(\rho)$ where $g(\rho) = C (1 - \beta \rho^n)$ is the density dependence with $C$ and $\beta$ being the constants of density dependence.

    Differentiating Eq.(27) with respect to $\rho$ one obtains equation for $X=0$:  
    
\begin{equation}
 \frac{\partial\epsilon}{\partial\rho} = [\frac{\hbar^2k_F^2}{5m\rho}] + \frac{J_{v00} C}{2} [1 - (n+1)\beta\rho^n] 
-\alpha J_{00} C [1 - \beta\rho^n]  [\frac{\hbar^2k_F^2}{10m}].
\label{eq28}
\end{equation}
\noindent
The saturation condition $\frac{\partial\epsilon}{\partial\rho} = 0$ at $\rho = \rho_{0}$, $\epsilon = \epsilon_{0}$, determines the equilibrium density of the cold SNM. Then for fixed values of the saturation energy per nucleon $\epsilon_0$ and the saturation density $\rho_{0}$ of the cold SNM, Eq.(27) and Eq.(28) with the saturation condition can be solved simultaneously to obtain the values of $\beta$ and $C$ which are given by 

\begin{equation}
 \beta = \frac{[(1-p)+(q-\frac{3q}{p})]\rho_{0}^{-n}}{[(3n+1)-(n+1)p+(q-\frac{3q}{p})]}
\label{eq29}
\end{equation} 
\noindent

\begin{equation}
 {\rm where}~~~~p = \frac{[10m\epsilon_0]}{[\hbar^2k_{F_0}^2]},~q=\frac{2\alpha\epsilon_0J_{00}}{J^0_{v00}}
\label{eq30}
\end{equation} 
\noindent
where $J^0_{v00} = J_{v00}(\epsilon^{kin}_0)$ which means $J_{v00}$ is evaluated at $\epsilon^{kin}=\epsilon^{kin}_0$, the kinetic energy part of the saturation energy per nucleon of SNM,  $k_{F_0} = [1.5\pi^2\rho_0]^{1/3}$ and 

\begin{equation}
 C = -\frac{[2\hbar^2k_{F_0}^2] }{ 5mJ^0_{v00} \rho_0 [1 - (n+1)\beta\rho_0^n -\frac{q\hbar^2k_{F_0}^2 (1-\beta\rho_0^n)}{10m\epsilon_0}]},
\label{eq31}
\end{equation} 
\noindent
respectively. Obviously, the constants $C$ and $\beta$ determined by this methodology depend upon $\epsilon_0$, $\rho_{0}$, the index $n$ of the density dependent part and through the volume integral $J^0_{v00}$, on the strengths of the M3Y interaction . 

    The calculations have been carried out by using the values of saturation density $\rho_0=0.1533$ fm$^{-3}$ \cite{Sa89} and saturation energy per nucleon $\epsilon_0=-15.26$ MeV \cite{CB06} for the SNM. $\epsilon_0$ is the co-efficient $a_v$ of the volume term of Bethe-Weizs\"acker mass formula, calculated by fitting the recent experimental and estimated Audi-Wapstra-Thibault atomic mass excesses \cite{Au03}. This term has been obtained by minimizing the mean square deviation incorporating correction for the electronic binding energy \cite{Lu03}. In a similar work, including surface symmetry energy term, Wigner term, shell correction and also the proton form factor correction to Coulomb energy, $a_v$ turns out to be 15.4496 MeV \cite{Ro06} ($a_v=14.8497$ MeV when $A^0$ and $A^{1/3}$ terms are also included). Using the standard values of $\alpha=0.005$ MeV$^{-1}$ for the parameter of energy dependence of the zero range potential and $n$=2/3, the values deduced for the constants $C$ and $\beta$ and the SNM incompressibility $K_\infty$ are, respectively, 2.2497, 1.5934 fm$^2$ and 274.7 MeV. The term $\epsilon_0$ is $a_v$ and its value of $-15.26\pm0.52$ MeV encompasses, more or less, the entire range of values. For this value of $a_v$ now the values of the constants of density dependence are $C=2.2497\pm0.0420$, $\beta=1.5934\pm0.0085$ fm$^2$ and the SNM incompressibility $K_\infty$ turns out to be $274.7\pm7.4$ MeV.  
    
\subsection{Symmetric and isospin asymmetric nuclear matter}

    The EoSs of the SNM and the IANM which describes energy per nucleon $\epsilon$ as a function of nucleonic density $\rho$ can be obtained by setting isospin asymmetry $X=0$ and $X \neq 0$, respectively, in Eq.(27). It is observed that the energy per nucleon $\epsilon$ for SNM is negative (bound) up to nucleonic density of $\sim2\rho_0$ while for pure neutron matter (PNM) $\epsilon>0$ and is always unbound by nuclear forces. 
    
    The compression modulus or incompressibility of the SNM, which is a measure of the curvature of an EoS at saturation density and is defined as $k_F^2\frac{\partial^2\epsilon}{\partial{k_F^2}} \mid_{k_F=k_{F_0}}$. It measures the stiffness of an EoS and can be theoretically obtained by using Eq.(27) for X=0. The IANM incompressibilities are evaluated at saturation densities $\rho_s$ with the condition of vanishing pressure which is $\frac{\partial\epsilon}{\partial\rho}|_{\rho=\rho_s}=0$. The incompressibility $K_s$ for IANM is therefore expressed as 

\begin{eqnarray}
 &&K_s = -\frac{3\hbar^2k_{F_s}^2}{5m} F(X) - \frac{9 J^s_v C n(n+1) \beta\rho_s^{n+1}}{2} \nonumber \\
 &&- 9\alpha J C [1-(n+1)\beta\rho_s^n] [\frac{\rho_s\hbar^2k_{F_s}^2}{5m}] F(X) \nonumber \\
 &&+ [\frac{3\rho_s\alpha J C (1-\beta\rho_s^n)\hbar^2k_{F_s}^2}{10m}] F(X), 
\label{eq32}
\end{eqnarray} 
\noindent
where $k_{F_s}$ implies that the $k_F$ is calculated at saturation density $\rho_s$. The term $J^s_v=J^s_{v00} + X^2 J^s_{v01}$ is $J_v$ evaluated at $\epsilon^{kin}=\epsilon^{kin}_s$ which is the kinetic energy part of the saturation energy per nucleon $\epsilon_s$ and $J=J_{00} + X^2 J_{01}$.

    In Table-I, IANM incompressibility $K_s$ as a function of $X$, for the standard value of $n=2/3$ and energy dependence parameter $\alpha=0.005$ MeV$^{-1}$, is provided. The magnitude of the IANM incompressibility $K_s$ decreases with $X$ due to lowering of the saturation densities $\rho_s$ with the isospin asymmetry $X$ as well as decrease in the EoS curvature. At high values of $X$, the IANM does not have a minimum which signify that it can never be bound by itself due to interaction of nuclear force. However, the $\beta$ equilibrated nuclear matter which is a highly neutron rich IANM exists in the core of the neutron stars since its energy per nucleon is lower than that of SNM at high densities. Although it is unbound by the nuclear interaction but can be bound due to very high gravitational field that can be realized inside neutron stars.      

\noindent 
\begin{table}
\centering
\caption{IANM incompressibility at different isospin asymmetry $X$ using the usual values of $n=\frac{2}{3}$ and  $\alpha=0.005$ MeV$^{-1}$.}
\begin{tabular}{ccc}
\hline
\hline
$X$&$\rho_s$& $K_s$      \\
\hline
 & fm$^{-3}$ &MeV    \\ 
\hline
 0.0&0.1533&274.69 \\ 
 0.1&0.1525&270.44 \\ 
 0.2&0.1500&257.68 \\ 
 0.3&0.1457&236.64 \\ 
 0.4&0.1392&207.62 \\ 
 0.5&0.1300&171.16 \\  
 0.6&0.1170&127.84 \\
 0.7&0.0980&78.38 \\
\hline
\hline
\label{table1}
\end{tabular} 
\end{table}

    It is worthwhile to mention that the RMF-NL3 incompressibility for SNM is 271.76 MeV \cite{La97,La99} which is very close to 274.7$\pm$7.4 MeV obtained from the present calculation. In spite of the fact that the parameters of the density dependence of DDM3Y interaction have been tuned to reproduce the saturation energy per nucleon $\epsilon_0$ and the saturation density $\rho_{0}$ of the cold SNM that are obtained from finite nuclei, the agreement of the present EoS with the experimental flow data, where the high density behavior looks phenomenologically confirmed, justifies its extrapolation to high density.

    The SNM incompressibility is experimentally determined from the compression modes isoscalar giant monopole resonance (ISGMR) and isoscalar giant dipole resonance (ISGDR) of nuclei. The violations of self consistency in HF-RPA calculations \cite{Sh06} of the strength functions of ISGMR and ISGDR cause shifts in the calculated values of the centroid energies. These shifts may be larger in magnitude than the current experimental uncertainties. In fact, due to the use of a not fully self-consistent calculations with Skyrme interactions \cite{Sh06}, the low values of $K_\infty$ in the range of 210-220 MeV were predicted. Skyrme parmetrizations of SLy4 type predict $K_\infty$ values lying in the range of 230-240 MeV \cite{Sh06} when this drawback is corrected. Besides that bona fide Skyrme forces can be built so that the $K_\infty$ for SNM is rather close to the relativistic value of $\sim$ 250-270 MeV. Conclusion may, therefore, be drawn from the ISGMR experimental data that the magnitude of $K_\infty \approx$ 240 $\pm$ 20 MeV.

    The lower values \cite{Lu04,Yo04} for $K_\infty$ are usually predicted by the ISGDR data. However, it is generally agreed upon that the extraction of $K_\infty$ in this case more problematic for various reasons. Particularly, for excitation energies \cite{Sh06} above 30 and 26 MeV for $^{116}$Sn and $^{208}$Pb, respectively, the maximum cross-section for ISGDR at high excitation energy decreases very strongly and can even fall below the range of current experimental sensitivity. The upper limit of the recent values \cite{Yo05} for the nuclear incompressibility $K_\infty$ for SNM extracted from experiments is rather close to the present non-relativistic mean field model estimate employing DDM3Y interaction which is also in agreement with the theoretical estimates of relativistic mean field (RMF) models. With Gogny effective interactions \cite{Bl80}, which include nuclei where pairing correlations are important, the results of microscopic calculations reproduce experimental data on heavy nuclei for $K_\infty$ in the range about 220 MeV \cite{Bl95}. It may, therefore, be concluded that the calculated value of 274.7$\pm$7.4 MeV is a good theoretical result and is only slightly too high compared to the recent acceptable value \cite{Vr03,Sh09} of $K_\infty$ for SNM in the range of 250-270 MeV.       
\begin{figure}[t]
\vspace{-0.0cm}
\eject\centerline{\epsfig{file=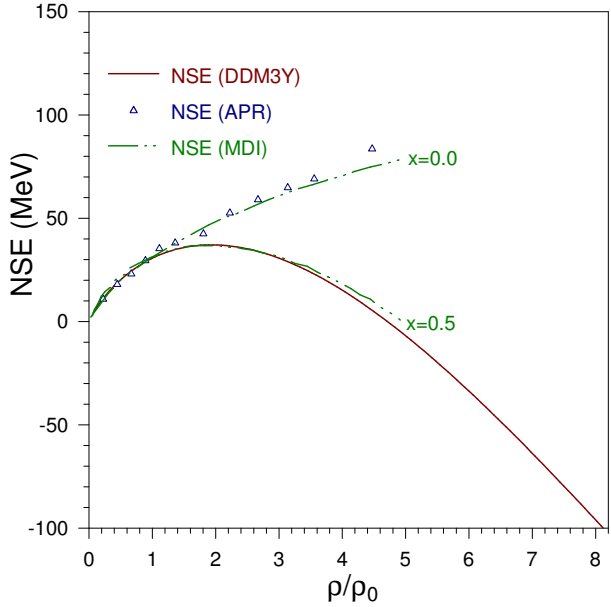,height=8cm,width=8cm}}
\caption{Plots of the nuclear symmetry energy NSE is as a function of $\rho/\rho_0$ for the present calculation using DDM3Y interaction and its comparison, with those for Akmal-Pandharipande-Ravenhall (APR) \cite{Ak98} and the MDI interactions for the variable x$=$0.0, 0.5 defined in Ref. \cite{Zh09}.}
\label{fig1}
\vspace{-0.0cm}
\end{figure}

\begin{figure}[t]
\vspace{0.0cm}
\eject\centerline{\epsfig{file=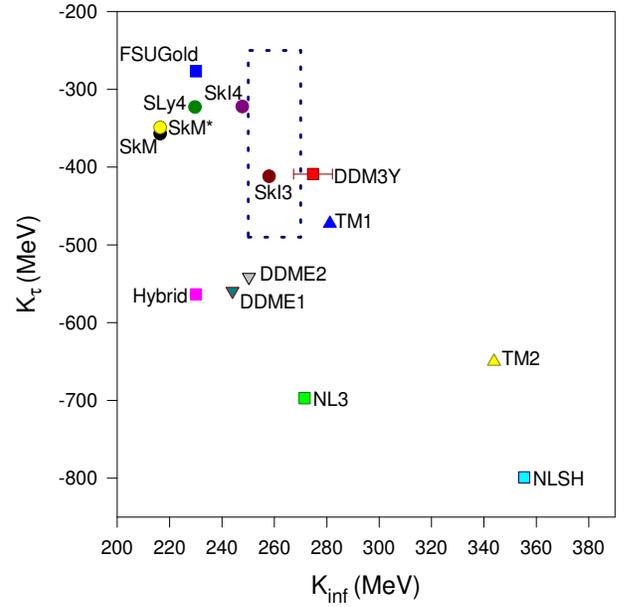,height=8cm,width=8cm}}
\caption{The plots of $K_\tau$ versus $K_\infty$ ($K_{inf}$). Present calculation (DDM3Y) is compared with other predictions as tabulated in Refs. \cite{Pi09,Sa07} and the dotted rectangular region encompasses the values of $K_\infty=250-270$ MeV \cite{Sh09} and $K_\tau=-370\pm120$ MeV \cite{Ch09}.}
\label{fig2}
\vspace{0.0cm}
\end{figure}

\subsection{Nuclear symmetry energy $\&$ its slope, incompressibility and isobaric incompressibility}

    The EoS of IANM, given by Eq.(27) can be expanded, in general, as

\begin{equation}
  \epsilon(\rho,X) =  \epsilon(\rho,0) + E_{sym}(\rho) X^2 + O ( X^4)
\label{eq33}
\end{equation} 
\noindent
where $E_{sym}(\rho)= \frac{1}{2} \frac{\partial^2\epsilon(\rho,X)}{\partial{X^2}} \mid_{X=0}$ is termed as the nuclear symmetry energy (NSE). The exchange symmetry between protons and neutrons in nuclear matter when one neglects the Coulomb interaction and assumes the charge symmetry of nuclear forces results in the absence of odd-order terms in $X$ in Eq.(33). To a good approximation, the density-dependent NSE $E_{sym}(\rho)$ can be obtained using the following equation \cite{Kl06}

\begin{equation}
 E_{sym}(\rho)=\epsilon(\rho,1) -\epsilon(\rho,0)
\label{eq34}
\end{equation}
\noindent
as the higher-order terms in $X$ are negligible. The above equation can be obtained using Eq.(27). It represents a penalty levied on the system as it departs from the symmetric limit of equal number of protons and neutrons. Thus, it can be defined as the energy required per nucleon to change the SNM to PNM. In Fig.-1 the plot of NSE as a function of $\rho/\rho_0$ is shown for the present calculation (DDM3Y) and compared with those for  Akmal-Pandharipande-Ravenhall \cite{Ak98} and MDI interactions \cite{Zh09}. 

    A constraint on the NSE at nuclear saturation density $E_{sym}(\rho_0)$ is provided by the volume symmetry energy coefficient $S_v$ which can be extracted from measured atomic mass excesses. The theoretical estimate for value of the NSE at saturation density $E_{sym}(\rho_0)$=30.71$\pm$0.26 MeV obtained from the present calculations (DDM3Y) is reasonably close to the value of $S_v$=30.048 $\pm$0.004 MeV extracted \cite{Mu07} from the measured atomic mass excesses of 2228 nuclei. The value of NSE at $\rho_0$ remains mostly the same which is 30.03$\pm$0.26 MeV if one uses the mathematical definition of $E_{sym}(\rho)= \frac{1}{2} \frac{\partial^2\epsilon(\rho,X)} {\partial{X^2}} \mid_{X=0}$ alternatively. The value of $E_{sym}(\rho_0)\approx$ 30 MeV \cite{St05,Da03,Po03} appears well established empirically. The different parameterizations of RMF models, which fit observables of isospin symmetric nuclei nicely, steers to a comparatively wide range of predictions of 24-40 MeV for $E_{sym}(\rho_0)$ theoretically. Our present result (DDM3Y) of 30.71$\pm$0.26 MeV is reasonably close to that obtained using Skyrme interaction SkMP (29.9 MeV) \cite{Be89}, Av18+$\delta v$+UIX$^*$ variational calculation (30.1 MeV) \cite{Ak98}. 
 
    The NSE $E_{sym}(\rho)$ can be expanded around the nuclear matter saturation density $\rho_0$ as 

\begin{equation}
 E_{sym}(\rho)= E_{sym}(\rho_0) + \frac{L}{3} {\Big (}\frac{\rho - \rho_0}{\rho_0}{\Big )}+ \frac{K_{sym}}{18}{\Big (}\frac{\rho - \rho_0}{\rho_0}{\Big )}^2
\label{eq35}
\end{equation}
\noindent
up to second order in density where $L$ and $K_{sym}$ represents the slope and curvature parameters of NSE at $\rho_0$ and hence $L= 3\rho_0 \frac{\partial E_{sym}(\rho)}{\partial\rho} \mid_{\rho=\rho_0}$ and $K_{sym}= 9\rho_0^2 \frac{\partial^2 E_{sym}(\rho)}{\partial {\rho^2}} \mid_{\rho=\rho_0}$. The $K_{sym}$ and $L$ highlights the density dependence of NSE around $\rho_0$ and carry important information at both high and low densities on the properties of NSE. Particularly, it is found that the slope parameter $L$ correlate linearly with neutron-skin thickness of heavy nuclei and it can be obtained from the measured thickness of neutron skin of heavy nuclei \cite{Pi05,Ch05,SL05}. Although there are large uncertainties in the experimental measurements, this has been possible \cite{Ce09} recently.

    Differentiation of Eq.(34) twice with respect to the nucleonic density $\rho$ using Eq.(27) provides \cite{BCM14}  
\begin{eqnarray}
 &&\frac{\partial E_{sym}}{\partial \rho} =\frac{2}{5}(2^{2/3}-1)\frac{E^0_F}{\rho}(\frac{\rho}{\rho_0})^{2/3}+\frac{C}{2}[1-(n+1)\beta\rho^n]  \nonumber \\
&&\times J_{v01}(\epsilon^{kin}_{X=1})-\frac{\alpha J_{01}C}{5}E^0_F(\frac{\rho}{\rho_0})^{2/3} [1-\beta\rho^n]F(1) \nonumber \\
&&-(2^{2/3}-1)\frac{\alpha J_{00}C}{5}E^0_F(\frac{\rho}{\rho_0})^{2/3}[1-\beta\rho^n] \nonumber \\ &&-\frac{3}{10}(2^{2/3}-1)\alpha J_{00}CE^0_F(\frac{\rho}{\rho_0})^{2/3}[1-(n+1)\beta\rho^n]
\label{eq36}
\end{eqnarray} 
\noindent

\begin{eqnarray}
&&\frac{\partial^2 E_{sym}}{\partial \rho^2} =-\frac{2}{15}(2^{2/3}-1)\frac{E^0_F}{\rho^2}(\frac{\rho}{\rho_0})^{2/3}-\frac{C}{2}n(n+1)\beta\rho^{n-1}\nonumber \\
&&\times J_{v01} (\epsilon^{kin}_{X=1}) -\frac{2\alpha J_{01}C}{5}\frac{E^0_F}{\rho}(\frac{\rho}{\rho_0})^{2/3}[1-(n+1)\beta\rho^n]F(1)  
 \nonumber \\
&&+\frac{\alpha J_{01}C}{15}\frac{E^0_F}{\rho}(\frac{\rho}{\rho_0})^{2/3}[1-\beta\rho^n]F(1) \nonumber \\
&&+(2^{2/3}-1)\frac{\alpha J_{00}C}{15}\frac{E^0_F}{\rho}(\frac{\rho}{\rho_0})^{2/3} [1-\beta\rho^n]  \nonumber \\
&&-\frac{2}{5}(2^{2/3}-1)\alpha J_{00}C\frac{E^0_F}{\rho}(\frac{\rho}{\rho_0})^{2/3} [1-(n+1) \beta\rho^n] \nonumber \\ 
&&+\frac{3}{10}(2^{2/3}-1)\alpha J_{00}CE^0_F(\frac{\rho}{\rho_0})^{2/3}n(n+1)\beta\rho^{n-1}.
\label{eq37}
\end{eqnarray} 
\noindent
Here the Fermi energy is $E^0_F=\frac{\hbar^2k_{F_0}^2}{2m}$ for the SNM at ground state and to evaluate the values of $L$ and $K_{sym}$, the definitions of which are provided after Eq.(35), along with Eqs.(36,37) at $\rho$=$\rho_0$ have been used. 

    The isobaric incompressibility $K_\infty (X)$ for infinite IANM can be expanded in the power series of isospin asymmetry parameter $X$ as $K_\infty (X) = K_\infty + K_\tau X^2 + K_4 X^4 + O(X^6)$. Compared to $K_\tau$ \cite{Ch09} the magnitude of the higher-order $K_4$ parameter is quite small in general. The former essentially characterizes the isospin dependence of the incompressibility at $\rho_0$ and expressed as $K_\tau= K_{sym}-6L- \frac{Q_0}{K_\infty}L =K_{asy}-\frac{Q_0}{K_\infty}L$ where the third-order derivative parameter of SNM at $\rho_0$ is $Q_0$, given by 
    
\begin{equation}
Q_0 = 27\rho_0^3 \frac{\partial^3 \epsilon(\rho,0)}{\partial {\rho^3}} \mid_{\rho=\rho_0}. 
\label{eq38}
\end{equation}
\noindent
One obtains, using Eq.(27), the following

\begin{eqnarray}
&&\frac{\partial^3 \epsilon(\rho,X)}{\partial {\rho^3}} 
=-\frac{CJ_v(\epsilon^{kin})n(n+1)(n-1)\beta\rho^{n-2}}{2}  \nonumber \\ &&+\frac{8}{45}\frac{E^0_F}{\rho^3}F(X)(\frac{\rho}{\rho_0})^\frac{2}{3} 
+\frac{3\alpha JC}{5}n(n+1)\beta\rho^{n-1}\frac{E^0_F}{\rho} \nonumber \\ 
&&\times F(X)(\frac{\rho}{\rho_0})^\frac{2}{3} 
+\frac{\alpha JC}{5}[1-(n+1)\beta\rho^n] \frac{E^0_F}{\rho^2}F(X)(\frac{\rho}{\rho_0})^\frac{2}{3} \nonumber \\
&&-\frac{4\alpha JC}{45}[1-\beta\rho^n]\frac{E^0_F}{\rho^2}F(X)(\frac{\rho}{\rho_0})^\frac{2}{3}
\label{eq39}
\end{eqnarray}
\noindent
where the Fermi energy $E^0_F$=$\frac{\hbar^2k_{F_0}^2}{2m}$ for the SNM at ground state,  $k_{F_0}$=$(1.5\pi^2\rho_0)^{\frac{1}{3}}$ and $J$=$J_{00}$+$X^2J_{01}$. Thus 

\begin{eqnarray}
&&\frac{\partial^3 \epsilon(\rho,0)}{\partial {\rho^3}} \mid_{\rho=\rho_0}
=-\frac{CJ_{v00}(\epsilon_0^{kin})n(n+1)(n-1)\beta\rho_0^{n-2}}{2} \nonumber \\  
&&+\frac{8}{45}\frac{E^0_F}{\rho_0^3}+\frac{3\alpha J_{00}C}{5}n(n+1)\beta\rho_0^{n-1} \frac{E^0_F}{\rho_0}+\frac{\alpha J_{00}C}{5} \nonumber \\   
&&\times [1-(n+1)\beta\rho_0^n]\frac{E^0_F}{\rho_0^2}-\frac{4\alpha J_{00}C}{45}[1-\beta\rho_0^n]\frac{E^0_F}{\rho_0^2}.
\label{eq40}
\end{eqnarray}
\noindent

\begin{figure}[t]
\vspace{0.0cm}
\eject\centerline{\epsfig{file=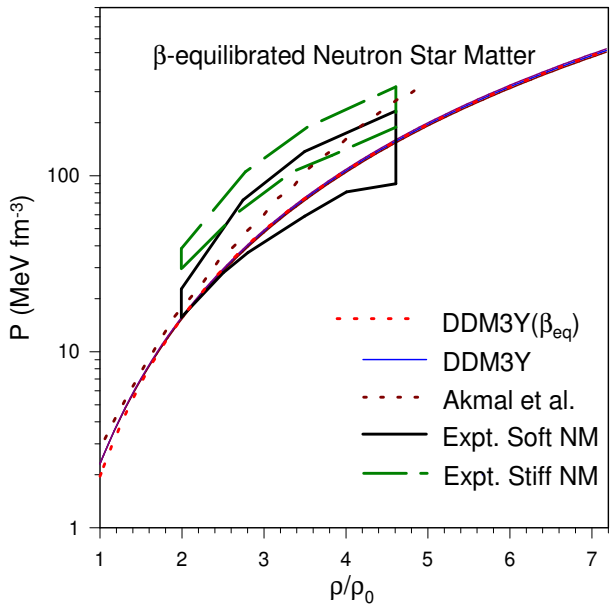,height=8cm,width=8cm}}
\caption{Plots for pressure P of dense nuclear matter as functions of $\rho/\rho_0$. The continuous line represents the pure neutron matter and the dashed line represents the $\beta$-equilibrated neutron star matter. The dotted line represents the same for the A18 model using variational chain summation of Akmal et al. \cite{Ak98}. The areas enclosed by the continuous and the dashed lines correspond to the pressure regions for neutron matter consistent with the experimental flow data after inclusion of the pressures from asymmetry terms with weak (soft NM) and strong (stiff NM) density dependences, respectively \cite{Da02}.}
\label{fig3}
\vspace{-0.14cm}
\end{figure}
\noindent 

\begin{figure}[t]
\vspace{0.0cm}
\eject\centerline{\epsfig{file=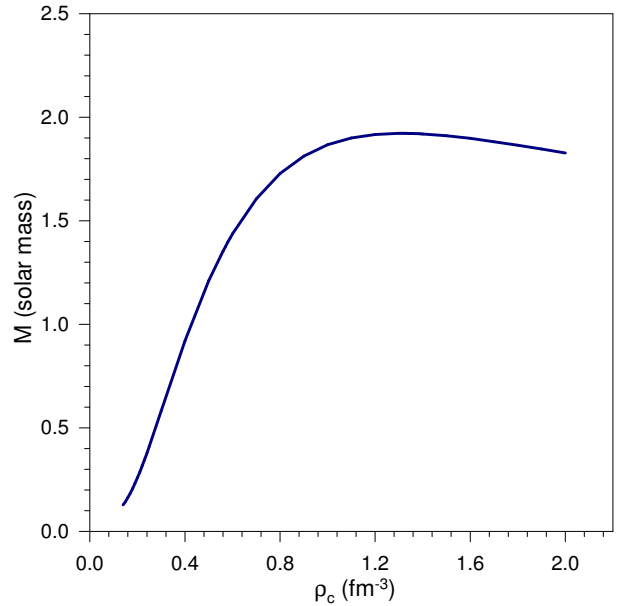,height=8cm,width=8cm}}
\caption
{Mass versus central baryonic density plot of slowly rotating neutron stars for the DDM3Y EoS. } 
\label{fig4}
\vspace{-0.14cm}
\end{figure}
\noindent

    For the calculations of $K_\infty$, $E_{sym}(\rho_0)$, $L$, $K_{sym}$ and $K_{\tau}$, the values of $\rho_0$=0.1533 fm$^{-3}$, $\epsilon_0=-15.26\pm0.52$ MeV for the SNM and $n=\frac{2}{3}$ \cite{CBS09} have been used. Using the improved quantum molecular dynamics transport model, the collisions involving $^{112}$Sn and $^{124}$Sn nuclei can be simulated to reproduce isospin diffusion data from two different observables and the ratios of proton and neutron spectra. The constraints on the density dependence of the NSE at subnormal density can be obtained \cite{Ts09} by comparing these data to calculations performed over a range of NSEs at $\rho_0$ and different representations of the density dependence of the NSE. The results for $K_\infty$, $L$, $E_{sym}(\rho_0)$ and density dependence of $E_{sym}(\rho)$ \cite{CBS09} of the present calculations are consistent with these constraints \cite{Ts09}. In Table-II, the values of $K_\infty$, $E_{sym}(\rho_0)$, $L$, $K_{sym}$ and $K_\tau$ are tabulated and compared with the corresponding quantities obtained from the RMF models \cite{Pi09}. 

    What is a reasonable value of incompressibility \cite{Sh06} remains controversial. In the following we present our results in the backdrop of others, without justifying any particular value for $K_\infty$, but for an objective view of the current situation which, we stress, is still progressing. In Fig.-2, the plot of $K_\tau$ versus $K_\infty$ for the present calculation (DDM3Y) has been compared with the predictions of SkI3, SkI4, SLy4, SkM, SkM*, FSUGold, NL3, Hybrid \cite{Pi09}, NLSH, TM1, TM2, DDME1 and DDME2 as given in Table-I of Ref. \cite{Sa07}. The recent values of $K_\infty=250-270$ MeV \cite{Sh09} and $K_\tau=-370\pm120$ MeV \cite{Ch09} are enclosed by the dotted rectangular region. Though both DDM3Y and SkI3 are within the above region, unlike DDM3Y the $L$ value for SkI3 is 100.49 MeV which is much above the acceptable limit of $58.9\pm16$ MeV \cite{Wa09,Ag13,Li13,Ag17}. Another recent review \cite{Kl17} also finds that $E_{sym}(\rho_0)=31.7\pm3.2$ MeV and $L=58.7\pm28.1$ MeV with an error for $L$ that is considerably larger than that for $E_{sym}(\rho_0)$. However, DDME2 is reasonably close to the rectangular region which has $L=51$ MeV. It is worthwhile to mention here that the DDM3Y interaction with the same ranges, strengths and density dependence that provides $L=45.11\pm0.02$, allows good descriptions of elastic and inelastic scattering, proton radioactivity \cite{BCS08} and $\alpha$ radioactivity of superheavy elements \cite{CSB06,scb07}. The present NSE increases initially with nucleonic density up to about 2$\rho_0$ and then decreases monotonically (hence `soft') and becomes negative at higher densities (about 4.7$\rho_0$) \cite{BCS08,CBS09} (hence `super-soft'). It is consistent with the recent evidence for a soft NSE at suprasaturation densities \cite{Zh09} and with the fact that the super-soft nuclear symmetry energy preferred by the FOPI/GSI experimental data on the $\pi^+/\pi^-$ ratio in relativistic heavy-ion reactions can readily keep neutron stars stable if the non-Newtonian gravity proposed in the grand unification theories is considered \cite{We09}.

\begin{table*}[htbp]
\centering
\caption{Comparison of the present results obtained using DDM3Y effective interaction with those of RMF models \cite{Pi09} for SNM incompressibility $K_\infty$, NSE at saturation density $E_{sym}(\rho_0)$, slope $L$ and the curvature $K_{sym}$ parameters of NSE, $K_{asy}$ and isobaric incompressibility $K_\tau$ of IANM  (all in MeV).}
\begin{tabular}{||c|c|c|c|c|c||}
\hline
\hline
Model&$K_\infty$&$E_{sym}(\rho_0)$&$L$&$K_{sym}$&$K_{asy}$\\ 
\hline
\hline
 This work &$274.7\pm7.4$&$30.71\pm0.26$&$45.11\pm0.02$&$-183.7\pm3.6$&$-454.4\pm3.5$\\
 &&&&$Q_0$=$-276.5\pm10.5$&$K_\tau$=$-408.97\pm3.01$ \\ 
\hline
 FSUGold&230.0&32.59&60.5&-51.3&-414.3\\
 &&&&$Q_0$=$-523.4$&$K_\tau$=$-276.77$\\
\hline
 NL3&271.5&37.29&118.2&+100.9&-608.3\\
 &&&&$Q_0$=$+204.2$&$K_\tau$=$-697.36$\\
\hline
 Hybrid&230.0&37.30&118.6&+110.9&-600.7\\ 
 &&&&$Q_0$=$-71.5$&$K_\tau$=$-563.86$\\
\hline
\hline
\end{tabular} 
\label{table2}
\end{table*}
\noindent 

\begin{table*}[htbp]
\centering
\caption{Results of the present calculations (DDM3Y) of symmetric nuclear matter incompressibility $K_\infty$, nuclear symmetry energy at saturation density $E_{sym}(\rho_0)$, the slope $L$ and  isospin dependent part $K_\tau$ of the isobaric incompressibility (all in MeV) \cite{CBS09} are tabulated along with the saturation density and the density, pressure and proton fraction at the core-crust transition for $\beta$-equilibrated neutron star matter.}
\begin{tabular}{||c|c|c|c||}
\hline
\hline
$K_\infty$&$E_{sym}(\rho_0)$&$L$&$K_\tau$ \\ 
\hline
 $274.7\pm7.4$&$30.71\pm0.26$&$45.11\pm0.02$&$-408.97\pm3.01$ \\ 
\hline
$\rho_0$&$\rho_t$& P$_t$ & x$_{p(t)}$ \\
\hline
0.1533 fm$^{-3}$&0.0938 fm$^{-3}$&0.5006 MeV fm$^{-3}$& 0.0308 \\
\hline
\hline
\end{tabular} 
\label{table3}
\end{table*}
\noindent 

\noindent     
\section{$\beta$-equilibrated neutron star matter} 
\label{Section 4}
    
    The $\beta$-equilibrated nuclear matter EoS is obtained by evaluating the asymmetric nuclear matter EoS at the isospin asymmetry $X$ determined from the $\beta$-equilibrium proton fraction $x_p =\frac{\rho_p}{\rho}$, obtained approximately by solving 

\begin{equation}
 \hbar c (3 \pi^2\rho x_p)^{1/3}= 4E_{sym}(\rho) (1 - 2 x_p).
\label{eq41}
\end{equation}
\noindent
The exact way of obtaining $\beta$-equilibrium proton fraction is by solving 

\begin{equation}
 \hbar c (3 \pi^2\rho x_p)^{1/3} = -\frac{\partial \epsilon(\rho,x_p)}{\partial x_p} = +2\frac{\partial \epsilon}{\partial X},
\label{eq42}
\end{equation}
\noindent
where isospin asymmetry $X=1-2x_p$.

    The pressure $P$ of PNM and $\beta$-equilibrated neutron star matter are plotted in Fig.-3 as functions of $\rho/\rho_0$. The continuous line represents the PNM and the dashed line (almost merges with the continuous line) represents the $\beta$-equilibrated neutron star matter (present calculations) whereas the dotted line represents the same using the A18 model using variational chain summation (VCS) of Akmal et al. \cite{Ak98} for the PNM. The areas enclosed by the continuous and the dashed lines in Fig.-3 correspond to the pressure regions for neutron matter consistent with the experimental flow data after inclusion of the pressures from asymmetry terms with weak (soft NM) and strong (stiff NM) density dependences, respectively \cite{Da02}. Although, the parameters of the density dependence of DDM3Y interaction have been tuned to reproduce $\rho_0$ and $\epsilon_0$ which are obtained from finite nuclei, the agreement of the present EoS with the experimental flow data, where the high density behaviour looks phenomenologically confirmed, justifies its extrapolation to high density. It is interesting to note that the RMF-NL3 incompressibility for SNM is 271.76 MeV \cite{La97,La99} which is about the same as 274.7$\pm$7.4 MeV obtained from the present calculation but the plot of $P$ versus $\rho/\rho_0$ for PNM of RMF using NL3 parameter set \cite{La97} does not pass through the pressure regions for neutron matter consistent with the experimental flow data \cite{Da02}. 
    
    The stability of the $\beta$-equilibrated dense matter in neutron stars is investigated and the location of the inner edge of their crusts and core-crust transition density and pressure are determined using the DDM3Y effective nucleon-nucleon interaction \cite{At14}. The stability of any single phase, also called intrinsic stability, is ensured by the convexity of $\epsilon(\rho,x_p)$. The thermodynamical inequalities allow us to express the requirement in terms of $V_{thermal}=\rho^2\Big[2\rho\frac{\partial \epsilon}{\partial \rho}+\rho^2\frac{\partial^2 \epsilon}{\partial \rho^2}-\rho^2\frac{(\frac{\partial^2 \epsilon}{{\partial\rho \partial x_p}})^2}{\frac{\partial^2 \epsilon}{\partial x_p^2}}\Big]$. The condition for core-crust transition is obtained by making $V_{thermal}=0$. The results for the transition density, pressure and proton fraction at the inner edge separating the liquid core from the solid crust of neutron stars are calculated and presented in Table-III for $n$=2/3. The symmetric nuclear matter incompressibility $K_\infty$, nuclear symmetry energy at saturation density $E_{sym}(\rho_0)$, the slope $L$ and  isospin dependent part $K_\tau$ of the isobaric incompressibility are also tabulated since these are all in excellent agreement with the constraints recently extracted from measured isotopic dependence of the giant monopole resonances in even-A Sn isotopes, from the neutron skin thickness of nuclei, and from analyses of experimental data on isospin diffusion and isotopic scaling in intermediate energy heavy-ion collisions.

\begin{figure}[t]
\vspace{0.0cm}
\eject\centerline{\epsfig{file=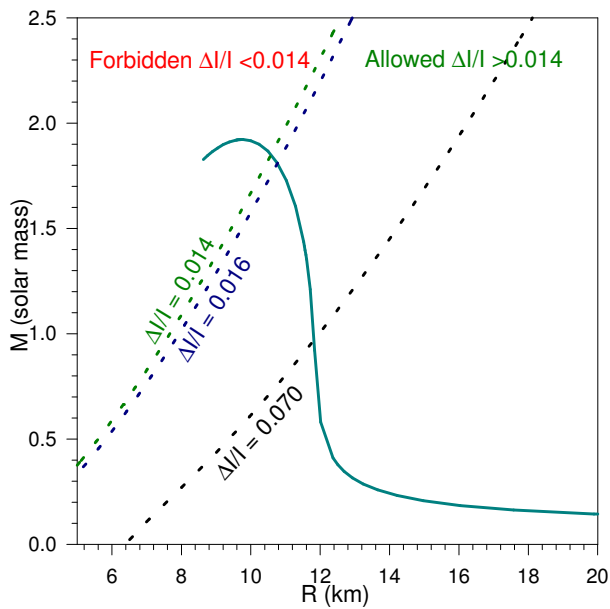,height=8cm,width=8cm}}
\caption{Mass-equatorial radius plot of slowly rotating neutron stars for the DDM3Y EoS. } 
\label{fig5}
\vspace{0.0cm}
\end{figure}
    
    The r-mode instability occurs when the gravitational-radiation driving time scale of the r-mode is shorter than the time scales of the various dissipation mechanisms that may occur in the interior of a neutron star. The nuclear EoS affects the time scales associated with the r-mode in two different ways viz. EoS defines the radial dependence of the mass density distribution which is the basic ingredient of the relevant integrals and defines the core-crust transition density and core radius which is the upper limit of these integrals. In Fig.-4, plot for mass versus central baryonic density of slowly rotating neutron stars is shown using the DDM3Y EoS for solving the Tolman-Oppenheimer-Volkoff Equation while in Fig.-5, the plot for mass-radius relation is shown. The mass-radius relation can be obtained for fixed values of $\frac{\Delta I}{I}$, the core-crust transition density $\rho_t$ and transition pressure P$_t$. This is then plotted in the same figure for $\frac{\Delta I}{I}$ equal to 0.014. For Vela pulsar, the constraint $\frac{\Delta I}{I}>1.4\%$ implies that allowed mass-radius lie to the right of the line defined by $\frac{\Delta I}{I} = 0.014$ (for $\rho_t$ = 0.0938 fm$^{-3}$ and P$_t$ = 0.5006 MeV fm$^{-3}$) \cite{At17}. The newer observational data \cite{Ho12} on Vela pulsar claims slightly higher estimate for $\frac{\Delta I}{I}>1.6\%$ based on glitch activity. This minute change neither affects the conclusions nor warrants any new idea of the neutron superfluidity extending partially into the core. However, if the phenomenon of crustal entrainment due to the Bragg reflection of unbound neutrons by the lattice ions is taken into account then \cite{An12,Ch13} a much higher fraction of the moment of inertia (7$\%$ instead of 1.4-1.6$\%$) has to be associated to the crust. The only reasonable constraint on mass of Vela pulsar is that it should exceed about one solar mass according to core-collapse supernova simulations. Therefore, the present calculations suggest that without entrainment, the crust is enough to explain the Vela glitch data and with entrainment, the crust is not enough since the mass of Vela pulsar would be below one solar mass (Fig.-5), in accordance with other studies \cite{An12,Ch13,Ho15,De16,Do16}.

    It is recently conjectured that there may be a good correlation between the core-crust transition density and the symmetry energy slope $L$ and it is predicted that this behaviour should not depend on the relation between $L$ and $K_\tau$ \cite{Du10}. On the contrary, no correlation of the transition pressure with $L$ was obtained \cite{Du10}. In Table-IV, variations of different quantities with parameter $n$ which controls the nuclear matter incompressibility are listed. It is worthwhile to mention here that the incompressibility increases with $n$. The standard value of $n$=2/3 used here has a unique importance because then the constant of density dependence $\beta$ has the dimension of cross section and can be interpreted as the isospin averaged effective nucleon-nucleon interaction cross section in ground state symmetric nuclear medium. For a nucleon in ground state nuclear matter $k_F\approx$ 1.3 fm$^{-1}$ and $q_0 \sim \hbar k_F c \approx$ 260 MeV and the present result for the `in medium' effective cross section is reasonably close to the value obtained from a rigorous Dirac-Brueckner-Hartree-Fock \cite{Sa06} calculations corresponding to such $k_F$ and $q_0$ values which is $\approx$ 12 mb. Using the value of $\beta$=1.5934 fm$^2$ along with the nucleonic density 0.1533 fm$^{-3}$, the value obtained for the nuclear mean free path $\lambda$ is about 4 fm which is in excellent agreement with that obtained using another method \cite{Si83}. Moreover, comparison of the theoretical values of symmetric nuclear matter incompressibility and isobaric incompressibility with the recent experimental values for $K_\infty=250-270$ MeV \cite{Sh09,St14} and $K_\tau=-370\pm120$ MeV \cite{Chen09} further justifies importance for our choice of $n$=2/3. It is interesting to mention here that the present EoS for $n$=2/3, provides the maximum mass for the static case is 1.9227 M$_\odot$ with radius $\sim$9.75 km and for the star rotating with Kepler's frequency it is 2.27 M$_\odot$ with equatorial radius $\sim$13.1 km \cite{Ch10}. However, for stars rotating with maximum frequency limited by the r-mode instability, the maximum mass turns out to be 1.95 (1.94) M$_\odot$ corresponding to rotational period of 1.5 (2.0) ms with radius about 9.9 (9.8) kilometers \cite{APP12} which reconcile with the recent observations of the massive compact stars $\sim$2 M$_\odot$ \cite{De10,An13}.   

\begin{table*}[htbp]
\centering
\caption{Variations of the core-crust transition density, pressure and proton fraction for $\beta$-equilibrated neutron star matter, symmetric nuclear matter incompressibility $K_\infty$, isospin dependent part $K_\tau$ of isobaric incompressibility, maximum neutron star mass, corresponding radius and crustal thickness with parameter $n$.}
\begin{tabular}{||c|c|c|c|c|c|c|c|c||}
\hline
\hline
$n$&$\rho_t$& P$_t$ & x$_{p(t)}$&$K_\infty$&$K_\tau$&Maximum NS mass & Radius& Crustal thickness \\
\hline
&&&&&&&& \\ 
1/6&0.0797 fm$^{-3}$&0.4134 MeV fm$^{-3}$& 0.0288&182.13 MeV&-293.42 MeV &1.4336 M$_\odot$&8.5671 km&0.4009 km \\ \hline
&&&&&&&&\\ 
1/3&0.0855 fm$^{-3}$&0.4520 MeV fm$^{-3}$& 0.0296&212.98 MeV&-332.16 MeV &1.6002 M$_\odot$&8.9572 km&0.3743 km \\ \hline
&&&&&&&&\\ 
1/2&0.0901 fm$^{-3}$&0.4801 MeV fm$^{-3}$& 0.0303&243.84 MeV&-370.65MeV &1.7634 M$_\odot$&9.3561 km&0.3515 km \\ \hline
&&&&&&&&\\ 
2/3&0.0938 fm$^{-3}$&0.5006 MeV fm$^{-3}$& 0.0308&274.69 MeV&-408.97 MeV &1.9227 M$_\odot$&9.7559 km&0.3318km \\ \hline
&&&&&&&&\\ 
1  &0.0995 fm$^{-3}$&0.5264 MeV fm$^{-3}$& 0.0316&336.40 MeV&-485.28 MeV &2.2335 M$_\odot$&10.6408 km&0.3088 km \\ \hline
\hline
\end{tabular}
\label{table4}
\vspace{3.5cm} 
\end{table*}
\noindent 

\noindent     
\section{Theoretical Calculations} 
\label{Section 5}

    The quantity which is of crucial importance in the evaluation of various times scales, as can be seen from Eq.(8) and Eq.(10), is the integral $\int_{0}^{R_c} \rho(r) r^6 dr$. This integral can be re-written in terms of energy density $\epsilon(r)=\rho(r)c^2$ and expressed in dimensionless form as 

\begin{equation}
I({R_{c}})=\int_{0}^{R_c} \left[\frac{\epsilon(r)}{\rm MeV fm^{-3}}\right]\left(\frac{r}{\rm km}\right)^6 d\left(\frac{r}{\rm km}\right)
\label{eq43}
\end{equation}
   
\begin{figure}[htbp]
\vspace{0.0cm}
\eject\centerline{\epsfig{file=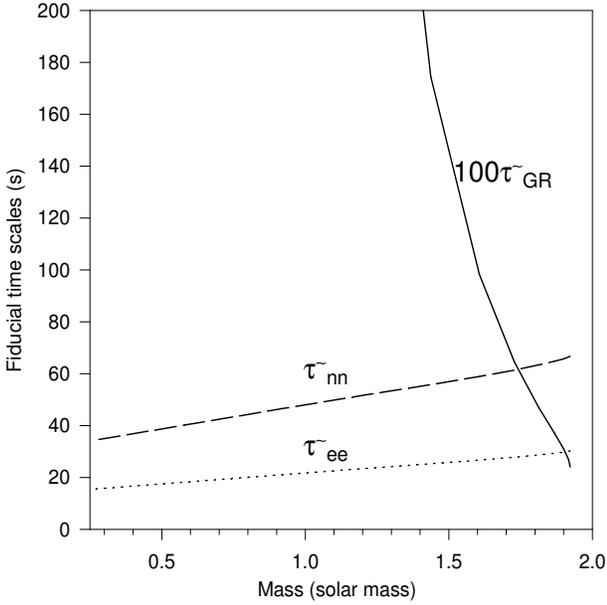,height=8cm,width=8cm}}
\caption{Plots of fiducial timescales with gravitational mass of neutron stars with DDM3Y EoS.} 
\label{fig6}
\vspace{0.0cm}
\end{figure}
    
\begin{figure}[htbp]
\vspace{0.0cm}
\eject\centerline{\epsfig{file=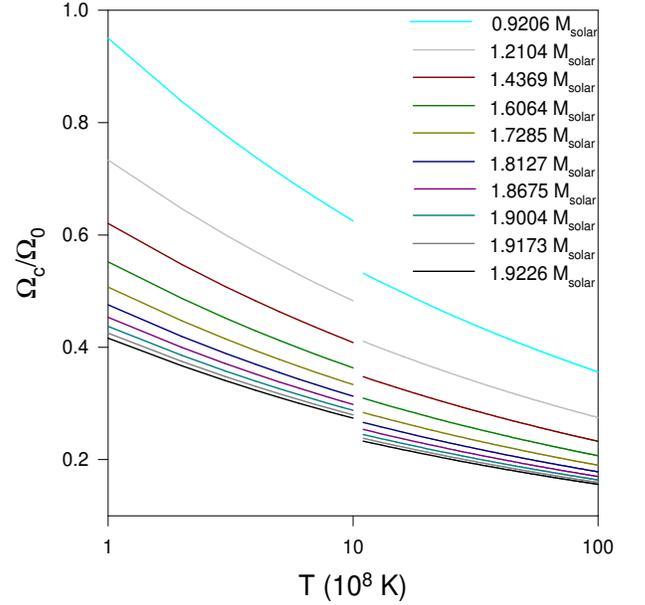,height=8cm,width=8cm}}
\caption{Plots of critical angular frequency with temperature for different masses of neutron stars.} 
\label{fig7}
\vspace{0.0cm}
\end{figure}

    The fiducial gravitational radiation timescale $\widetilde{\tau}_{GR}$ from Eq.(8) and Eq.(15), is given by
\begin{equation}
\widetilde{\tau}_{GR}=-0.7429\left[\frac{R}{\rm  km}\right]^{9} \left[\frac{1M_{\odot}}{M}\right]^{3}\left[I(R_c)\right]^{-1} ({\rm s})
\label{eq44}
\end{equation}
where $R$ and $r$ are in km and $M$ in $M_{\odot}$.

    The fiducial shear viscous timescale $\widetilde{\tau}_{SV}$ for electron-electron scattering and neutron-neutron scattering can be obtained from Eq.(10), Eqs.(12-14) as 
    
\begin{eqnarray}
\widetilde{\tau}_{ee}=0.1446\times10^8\left[\frac{R}{\rm km}\right]^{3/4}\left[\frac{1M_{\odot}}{M}\right]^{1/4} \left[\frac{\rm km}{R_c}\right]^{6}\nonumber \\
\times\left[\frac{\rm g~cm^{-3}}{\rho_t}\right]^{1/2}\left[\frac{\rm MeV fm^{-3}}{\epsilon_t}\right] \left[I(R_c)\right] ({\rm s})
\label{eq45}
\end{eqnarray}
\begin{eqnarray}
\widetilde{\tau}_{nn}=19\times10^8\left[\frac{R}{\rm km}\right]^{3/4}\left[\frac{1M_{\odot}}{M}\right]^{1/4}\left[\frac{\rm km}{R_c}\right]^{6}\nonumber \\
\times\left[\frac{\rm g~cm^{-3}}{\rho_t}\right]^{5/8}\left[\frac{\rm MeV fm^{-3}}{\epsilon_t}\right] \left[I(R_c)\right] ({\rm s})
\label{eq46}
\end{eqnarray}
where the transition density $\rho_t$ is expressed in g cm$^{-3}$ and $\epsilon_t$ is the energy density expressed in MeV fm$^{-3}$ at transition density.
 
\begin{figure}[t]
\vspace{0.0cm}
\eject\centerline{\epsfig{file=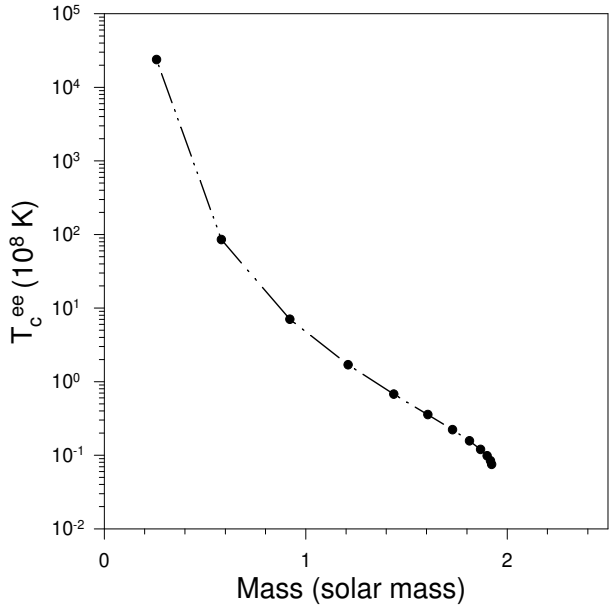,height=8cm,width=8cm}}
\caption{Plots of critical temperature versus mass.} 
\label{fig8}
\vspace{0.2cm}
\end{figure}

\noindent
\section{Results and discussion}
\label{Section 6}

    In Fig.-6 plots of the fiducial timescales with the gravitational masses of neutron stars are shown for the DDM3Y EoS. It is seen that the gravitational radiation timescale falls rapidly with increasing mass while the viscous damping timescales increase approximately linearly.

    By knowing the fiducial gravitational radiation and shear viscous timescales, the temperature T dependence of the critical angular velocity $\Omega_c$ of the r-mode $(l=2)$ can be studied from Eq.(17). In Fig.-7, $\frac{\Omega_c}{\Omega_0}$ is shown as a function of temperature T for several masses of neutron stars for the DDM3Y EoS. The plots act as boundaries of the r-mode instability windows. Neutron stars lying above the plots (whose angular frequency is greater than the critical frequency) possess unstable r-modes and hence emit gravitational waves, thus reducing their angular frequencies. Once their angular velocities reach the critical frequency they enter the region below the plots, where the r-modes become stable and hence stop emitting gravitational radiation. In computing the instability windows in Fig.-7, the fiducial shear viscous timescale $\widetilde \tau_{ee}$ given in Eq.(45) is substituted for $\widetilde \tau_{SV}$ in Eq.(17) for temperatures T $\leq$ $10^9$ K and $\tau_{nn}$ from Eq.(46) is used for T $>$ $10^9$ K.
 
    Fig.-8 depicts the plot of the critical temperature as a function of mass. The electron-electron scattering shear viscosity timescale is used for the calculation of $T_c$. We see that the critical temperature rapidly decreases with mass. The explanation is straightforward. From Fig.-7 we see that for fixed $T$, $\frac{\Omega_c}{\Omega_0}$ rapidly decreases with increasing mass. Since $T=T_c$ when $\Omega_c=\Omega_K$ and $\Omega_K$ rapidly increases with mass and hence $T_c$ falls, vide Eq.(19).  
 
\begin{figure}[t]
\vspace{0.0cm}
\eject\centerline{\epsfig{file=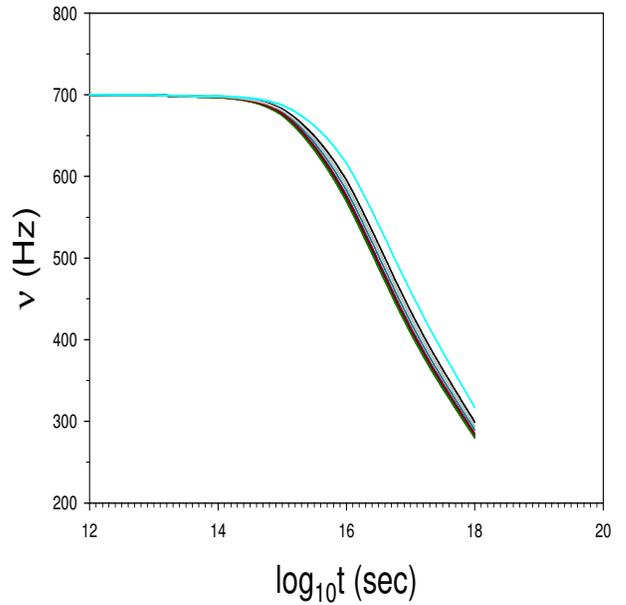,height=8cm,width=8cm}}
\caption{Plots of time evolution of frequencies.} 
\label{fig9}
\vspace{0.0cm}
\end{figure}
          
    From Fig.-7 and Fig.-8 we see that the critical frequency and critical temperature decrease with mass and hence the r-mode instability window increases with the same. This means that for the same EoS and temperature, the massive configurations are more probable to r-mode instability and hence emission of gravitational waves than the less massive ones. This can be indirectly inferred from Fig.-6 where $\widetilde \tau_{GR}$ is much less than $\widetilde \tau_{ee}$ and $\widetilde \tau_{nn}$ for massive neutron stars and vice-versa for low mass neutron stars. Hence isolated young massive neutron stars have high probability for emission of gravitational waves through r-mode instability.  
 
    It is worth noting that $\Omega_c$ is dependent on the density dependence of the symmetry energy and thus on $L$. Again, $R$, $R_c$, $I(R_c)$ and $\rho_t$ depend on $L$. Hence, for a fixed mass and temperature, $\Omega_c$ is dependent on the above parameters via the relation,
    
\begin{equation}
\Omega_c \sim \frac{R_c^{12/11}}{[I(R_c)]^{4/11}}\rho_t^{3/11}
\label{eq47}
\end{equation} 
In our case $L$, $\rho_t$ and $R_c$ are constants for a fixed neutron star mass and temperature.
    As a neutron star enters into the instability region due to accretion of mass from its companion, the amplitude of the r-mode $\alpha_r$ increases till reaching a saturation value. At this point the neutron star emits gravitational wave and releases its angular momentum and energy and spins down to the region of stability. Using the ideal condition that the decrease in temperature due to emission of gravitational wave is compensated by the heat produced due to viscous effects, the time evolution of spin angular velocity and spin down rate can be calculated for a neutron star from Eq.(23) and Eq.(25), respectively, provided M, T, $\Omega_{in}$ and $\alpha_r$ of the star is known. For the schematic values $\nu_{in}=\frac{\Omega_{in}}{2\pi}=700$ Hz and $\alpha_r=2 \times 10^{-7}$ used by Moustakidis \cite{Moustakidis2015}, the evolutions of spin are calculated for various neutron star masses and shown in Fig.-9. In Fig.-10 the spin down rates has been shown for these masses. In Fig.-11 the spin down rates as functions of spin frequency are shown.
    
    Some mention is to be made about the dependency of the critical frequency $\Omega_c$ on the symmetry energy slope parameter $L$. Although the slope $L$ depends on the strengths and ranges of the Yukawas for the DDM3Y EoS, it does not depend on the power of the density dependence $n$ and has a constant value of 45.1066 MeV. In a recent work, the critical frequency as a function of $L$ of the pulsar 4U 1608-52 was plotted using an estimated core temperature $\sim 4.55\times 10^8$ K and with different models of the EoS. In accordance with the Fig.-6 of \cite{Vidana2012}, using the measured spin frequency and the estimated core temperature, if the mass of 4U 1608-52 is $1.4M_\odot$ then it should marginally be unstable ($\Omega_c$ is smaller than its spin frequency), since the radius obtained from our mass-radius relation (Fig.-5) is $\sim 11.55$ kms and higher than 11.5 kms. In case of the highest mass configuration of 1.9227 M$_\odot$ with a radius of $\sim$9.75 kms, it is also likely to be in the instability region as $L<50$ MeV for our EoS. Thus we stress the fact that the r-mode instability window is enlarged for isolated neutron stars with a rigid crust if we consider the dissipation to be at the crust-core interface in agreement with \cite{We12}.          

\begin{figure}[t]
\vspace{0.0cm}
\eject\centerline{\epsfig{file=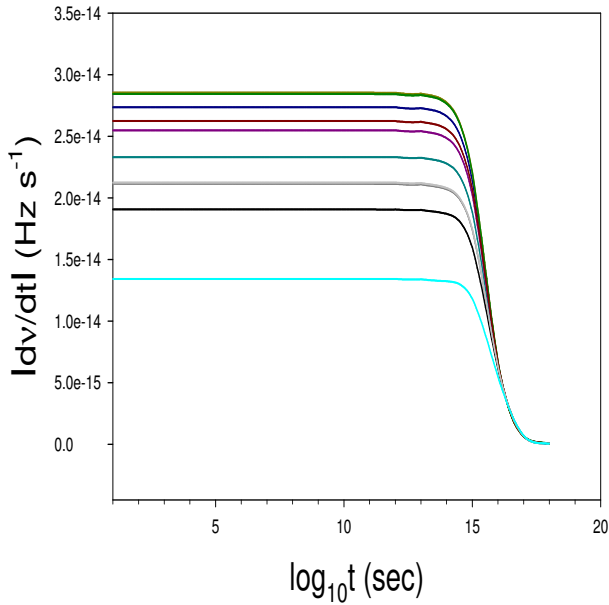,height=8cm,width=8cm}}
\caption{Plots of time evolution of spin-down rates.} 
\label{fig10}
\vspace{0.0cm}
\end{figure}

\begin{figure}[t]
\vspace{0.0cm}
\eject\centerline{\epsfig{file=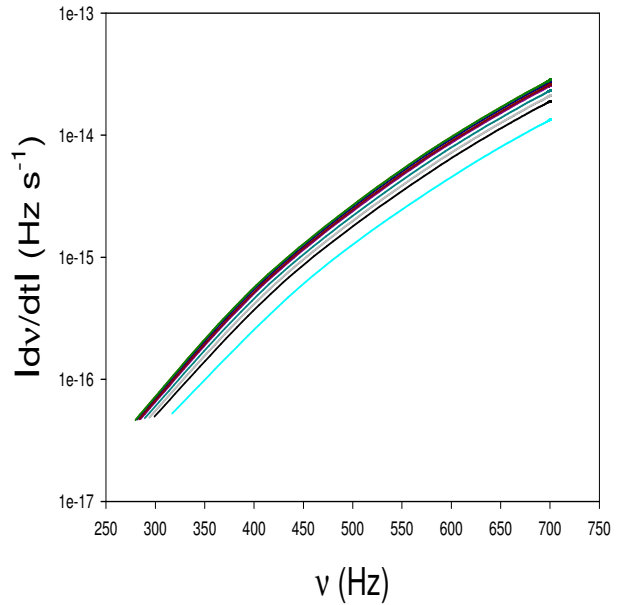,height=8cm,width=8cm}}
\caption{Plots of spin-down rates versus frequencies.} 
\label{fig11}
\vspace{-0.07cm}
\end{figure}

    The calculations are performed for five different $n$ values that correspond to SNM incompressibility ranging from $\sim$180-330 MeV. For each case, the constants $C$ and $\beta$ obtained by reproducing the ground state properties of SNM become different leading to five different sets of these three parameters. We certainly can not change strengths and ranges of the M3Y interaction. In Table-IV, the variations of the core-crust transition density, pressure and proton fraction for $\beta$-equilibrated neutron star matter, symmetric nuclear matter incompressibility $K_\infty$, isospin dependent part $K_\tau$ of isobaric incompressibility, neutron star's maximum mass with corresponding radius and crustal thickness with parameter $n$ are listed. It is important to mention here that recent observations of the binary millisecond pulsar J1614-2230 by P. B. Demorest et al. \cite{De10} suggest that the masses lie within $1.97\pm0.04$ M$_\odot$. Recently the radio timing measurements of the pulsar PSR J0348 + 0432 and its white dwarf companion have confirmed the mass of the pulsar to be in the range 1.97-2.05 M$_\odot$ at 68.27$\%$ or 1.90-2.18 M$_\odot$ at 99.73$\%$ confidence \cite{An13}. The observed $1.97\pm0.04$ M$_\odot$ neutron star rotates with 3.1 ms and results quoted in Table-5.2 are for non-rotating case. For rotating stars \cite{BCM14} present EoS predict masses higher than the lower limit of 1.93 M$_\odot$ for maximum mass of neutron stars. We used the same value of $\rho_0=0.1533$ fm$^{-3}$ since we wanted to keep consistency with all our previous works on nuclear matter. We would like to mention that if instead we would have used the value of 0.16 fm$^{-3}$ for $\rho_0$, the value of $K_\infty$ would have been slightly higher by $\sim$2 MeV and correspondingly maximum mass of neutron stars by $\sim$ 0.01 M$_\odot$. 

\noindent
\section{Summary and conclusions}
\label{Section 7}

    In the present work we have studied the r-mode instability of slowly rotating neutron stars with rigid crusts with their EoS obtained from the DDM3Y effective nucleon-nucleon interaction. This EoS provides good descriptions for proton, $\alpha$ and cluster radioactivities, elastic and inelastic scattering, symmetric and isospin asymmetric nuclear matter and neutron star masses and radii \cite{BCM14}. We have calculated the fiducial gravitational radiation and shear viscosity timescales within the DDM3Y framework for a wide range of neutron star masses. It is observed that the gravitational radiation timescale decreases rapidly with increasing neutron star mass while the viscous damping timescales exhibit an approximate linear increase. Next, we have studied the temperature dependence of the critical angular frequency for different neutron star masses. The implication is that for neutron stars rotating with frequencies greater than their corresponding critical frequencies have unstable r-modes leading to the emission of gravitational waves. Further, our study of the variation of the critical temperature as a function of mass shows that both the critical frequency and temperature decrease with increasing mass. The conclusion is that massive hot neutron stars are more susceptible to r-mode instability through gravitational radiation. Finally we have calculated the spin down rates and angular frequency evolution of the neutron stars through r-mode instability. We have also pointed out the fact that the critical frequency depends on the EoS through the radius and the symmetry energy slope parameter $L$. If the dissipation of r-modes from shear viscosity acts along the boundary layer of the crust-core interface then the r-mode instability region is enlarged to lower values of $L$.

\end{document}